\documentclass[twocolumn,showpacs,showkeys,prd,nofootinbib,amsmath,amssymb]{revtex4}

\usepackage{pgf,tikz}
\usepackage{mathrsfs}
\usetikzlibrary{arrows}

\usepackage{color}
\usepackage{graphicx}
\begin{document}

\def\half{\textstyle{1\over2}}
\def\third{\textstyle{1\over3}}
\def\quarter{\textstyle{1\over4}}
\newcommand{\beq}{\begin{eqnarray}}
\newcommand{\eeq}{\end{eqnarray}}
\newcommand{\be}{\begin{equation}}
\newcommand{\ee}{\end{equation}}
\newcommand{\bea}{\begin{eqnarray}}
\newcommand{\eea}{\end{eqnarray}}
\newcommand{\ud}{\mathrm{d}}
\newcommand{\GN}{\!}
\newcommand{\ba}{\begin{eqnarray}}
\newcommand{\ea}{\end{eqnarray}}

\title{Strong Deflection Limit Lensing  Effects  in the Minimal Geometric Deformation and Casadio-Fabbri-Mazzacurati Solutions}

\author{R.~T.~Cavalcanti}
\email{rogerio.cavalcanti@ufabc.edu.br}
 \affiliation{
Centro de Ci\^encias Naturais e Humanas, 
Universidade Federal do ABC,\\ 09210-580, Santo Andr\'e - SP, Brazil}

\author{A. Goncalves da Silva}
\email{allan.goncalves@ufabc.edu.br}
 \affiliation{
Centro de Ci\^encias Naturais e Humanas, 
Universidade Federal do ABC,\\ 09210-580, Santo Andr\'e - SP, Brazil}

\author{Rold\~ao~da~Rocha}
\email{roldao.rocha@ufabc.edu.br}
\affiliation{Centro de   Matem\'atica, Computa\c c\~ao e Cogni\c c\~ao, 
Universidade Federal do ABC,\\ 09210-580, Santo Andr\'e - SP, Brazil.}


\date{\today}

\begin{abstract}
In this paper we apply the strong deflection limit approach to investigate the
gravitational lensing phenomena beyond general relativity. {}{This} is accomplished by considering the lensing effects related to black hole solutions {}{that emerge} out of  the domain of Einstein gravity, namely, the ones acquired from the method of geometric deformation and the Casadio-Fabbri-Mazzacurati brane-world black holes.  The lensing
observables, for those brane-world black hole metrics, are compared with the {}{standard}  ones for the
Schwarzschild case. We prove that brane-world black holes could
have significantly different observational signatures, compared to the Schwarzschild
black hole, with terms containing the post-Newtonian parameter, for the case of the Casadio-Fabbri-Mazzacurati, and {}{terms with} variable brane-world tension, for the method of geometric deformation.
\end{abstract}



\maketitle

\section{Introduction} \label{sec:intro}
Black hole solutions of the Einstein equations in general relativity (GR) are useful tools for investigating the spacetime structure, the collapse of compact stellar distributions, and quantum effects in  theories of gravity.
In particular, extended models of GR ~\cite{maartens} lead to important consequences, not only to black hole
physics~\cite{Dimopoulos:2001hw}, but also to particle physics, cosmology, and to the astrophysics
of supermassive objects~\cite{daRocha:2013ki}. Hence, exploring the gravitational phenomena, in which these consequences come to light,  provides a way to pave extended models, as well as to test the limits of GR. The gravitational lensing represents one of those phenomena. The deflection of light by a gravitational potential was first observed in 1919 by Dyson, Eddington and Davidson \cite{Dyson:1920cwa}, whose modern refinements  has  become one of the experimental grounds of GR. Thereafter, the deflection of light was found to imply  lens effects, that could magnify or {}{even} create multiple images of astrophysical objects \cite{Zwicky:1937zzb}. This was a landmark for a contemporary field of research, known as gravitational lensing (GL). The works of Liebes  \cite{Liebes:1964zz} and Refsdal \cite{Refsdal:1964yk} developed the theory of gravitational lensing in the so-called weak field limit, where the lens equations and the expression for the deflection angle are quite simplified, hence allowing one to solve them exactly. The predictions of the theory, in this regime, have been thoroughly supported by experiments {}{and observations}. It is worth to mention, for example,  the observation of twin quasars separated by arc-seconds ({}{arcsec}), however at same redshifts and magnitudes \cite{Walsh:1979nx}, images of distorted galaxies inside another galaxy \cite{paczynski1987giant}, among others \cite{deXivry:2009ci}. With all this success in the weak field limit, the question about what happens in the strong field regime was driven by the possibility of existence of galactic supermassive black holes, at the centre of our galaxy. The subject was brought back by Virbhadra and Ellis \cite{Virbhadra:1999nm}, that theoretically investigated the strong field region of a Schwarzschild type lens. They found that, in the strong field regime, the theory predicts a large number of images of an observed  object -- theoretically, an infinite sequence of images, with an adherent point. The result contrasts with a pair of images, or an Einstein ring, predicted in the weak field limit regime. Following the work of Virbhadra and Ellis, Bozza \cite{Bozza:2001xd,Bozza:2002zj} has found an interesting simplification for the lens equation in such regime, finding the expression for observables quantities in the so-called strong deflection limit  (SDL) regime. Bozza proved  that, when the angle between the source and the lens tends to zero, the deflection angle diverges logarithmically. Furthermore, it can be integrated up to first order, wherefrom the GL observables can be derived.

 In a series of papers,  Keeton and Petters studied a general framework for the weak field limit, comprising the case of a static and spherically symmetric solution \cite{Keeton:2005jd}, the post-Newtonian metrics \cite{Keeton:2006sa} and brane-world gravity \cite{Keeton:2006di}. Whisker considered the gravitational lensing at strong deflection limit \cite{Whisker:2004gq}, as a way to seek for signatures of solutions of five-dimensional (5D) brane-world gravity. For some 5D brane-world solutions, the difference in the observables were found to be very small from the four-dimensional (4D) Schwarzschild  case. Thereafter,  Bozza revised the theoretical and observational aspects of gravitational lensing produced by black holes \cite{Bozza:2009yw}. {}{In a} recent paper, the SDL approach was applied to study Galileon black holes \cite{Zhao:2016kft}.

 In this paper, we apply the SDL to obtain the GL observables for two remarkable black holes solutions beyond GR, namely the Casadio-Fabbri-Mazzacurati  and the minimal geometric deformation. Solutions of the {}{effective} Einstein field equations in brane-world models,  are not, in general, univocally governed  by the matter stress-tensor. Indeed, gravity can propagate into the bulk, hence generating a Weyl term on the brane, accordingly \cite{maartens}.
The Casadio-Fabbri-Mazzacurati (CFM)  metrics generalise the Schwarzschild solution and have a parametrised post-Newtonian
(PPN) {}{parameter, that is  observationally -- by the Cassini probe -- and experimentally -- by very-long-baseline interferometry -- bounded to be $\delta\approx 1$ \cite{WILL}}. The PPN measures, in particular, the difference between
 the inertial and the gravitational masses and, furthermore, affects perihelia shifts
and accounts Nordtvedt effect, as moreover constrained by the Lunar Laser Ranging Experiment  (LLRE) \cite{Casadio:2002uv,cfabbri,WILL}. 
The PPN $\delta$ allows comparing experimental data in the weak-field limit, which is adequately accurate, in order  to include solar system tests~\cite{WILL}. 
In fact, the parameter $\delta$ is the usual Eddington-Robertson-Schiff parameter used to describe
the classical tests of GR. {The minimal geometric deformation (MGD), on the other hand, was introduced in the context of investigating the outer spacetime
around (spherically symmetric) self-gravitating systems. It includes stars or similar compact astrophysical objects \cite{ovalle2007,Casadio:2013uma,Casadio:2015jva,Casadio:2015gea,covalle2} likewise,
in Randall-Sundrum-like brane-world setups. To solve the brane effective Einstein field
equations is an intricate endeavour, whose analytical solutions are scarce in the literature \cite{maartens}. The MGD naturally encompasses the (variable) brane tension, and has been employed to derive exact, physically feasible, solutions for spherically symmetric, for non-uniform, and inner stellar distributions, to generate other physical inner stellar solutions, and their microscopic counterparts, accordingly. MGD
 can be thought of being induced by non-local outer Weyl
stresses that are evinced from bulk gravitons. 
In addition, the deformation of the time component  of the metric has been recently accomplished, thus characterising the so-called extended MGD procedure \cite{Casadio:2015gea}. The MGD further encompasses tidally charged metrics of extremal black hole with degenerate horizons, outer  solutions for self-gravitating systems induced by 5D Weyl fluids -- which can represent a Weyl atmosphere, accordingly \cite{ovalle2007,Ovalle:2007bn,Casadio:2015gea}. 

Our aim in this paper is to 
derive and analyse the gravitational lensing effects regarding the CFM and MGD solutions.
The paper is organised as follows. In Sect. II we introduce the fundamental   equations and {}{the  physical framework  underlying the} gravitational lensing, mainly focusing on the SDL regime. In Sect. III,  we briefly revisit the MGD as well as the CFM solutions, in order to fix the notation and to 
define the main setup that shall be further analysed. Sect. IV is devoted to the observable quantities in the SDL regime, containing  our results and their  analysis, for the solutions discussed in the Sect. III. To conclude, in Sect. V we summarise the results, point out the  concluding remarks and perspectives, and discuss the possibility of detection/observation  of signatures of the MGD and CFM solutions. 

\section{Gravitational lensing and the {strong deflection limit} (SDL)}
The correct prediction of the light deflection  angle by the Sun is one  of the greatest and earliest achievements of GR. Gravitational lensing regards the deflection of light through a gravitational field, being related with, for instance, the Hubble constant and {}{the} cosmological constant, among others, arising as an important tool to probe the physical properties of astrophysical objects. The general setup of a gravitational lens is depicted in Fig. 1. The light of a source \textit{S} is deflected, when it passes {}{through} a lens \textit{L}. The image of \textit{S} appears to the observer \textit{O}, in a position characterised by the angle $\theta$, instead of the angle $\beta$, and the deflection angle $\alpha$ \cite{Carroll:2004st}. The lens equation \footnote{{Eq. \eqref{tnnn} is valid only if the light rays emitted from the source are coplanar to the lens at $r_{0}$.  Details on this approximation can be found in \cite{Bozza:2008ev}.} }, relating  these angles, reads \begin{equation}
\tan\beta = \tan\theta - \frac{D_{ls}}{D_{os}}\left[\tan\theta - \tan\left(\theta-\alpha \right)\right].\label{tnnn}
\end{equation}  
Hence, an object position $\beta$ provides the angle $\alpha$, in order to solve Eq. (\ref{tnnn}) for the image position $\theta$. Basic assumptions are that the lens effect is induced by local matter inhomogeneities and the source and the observer are in an asymptotic  flat spacetime. The integration of the null geodesic \cite{weinberg1972gravitation},  associated with the angular deflection as a function of radial distance $r$, in the vicinity of a metric
\begin{figure}[!htb]
\centering
\includegraphics[scale=.99]{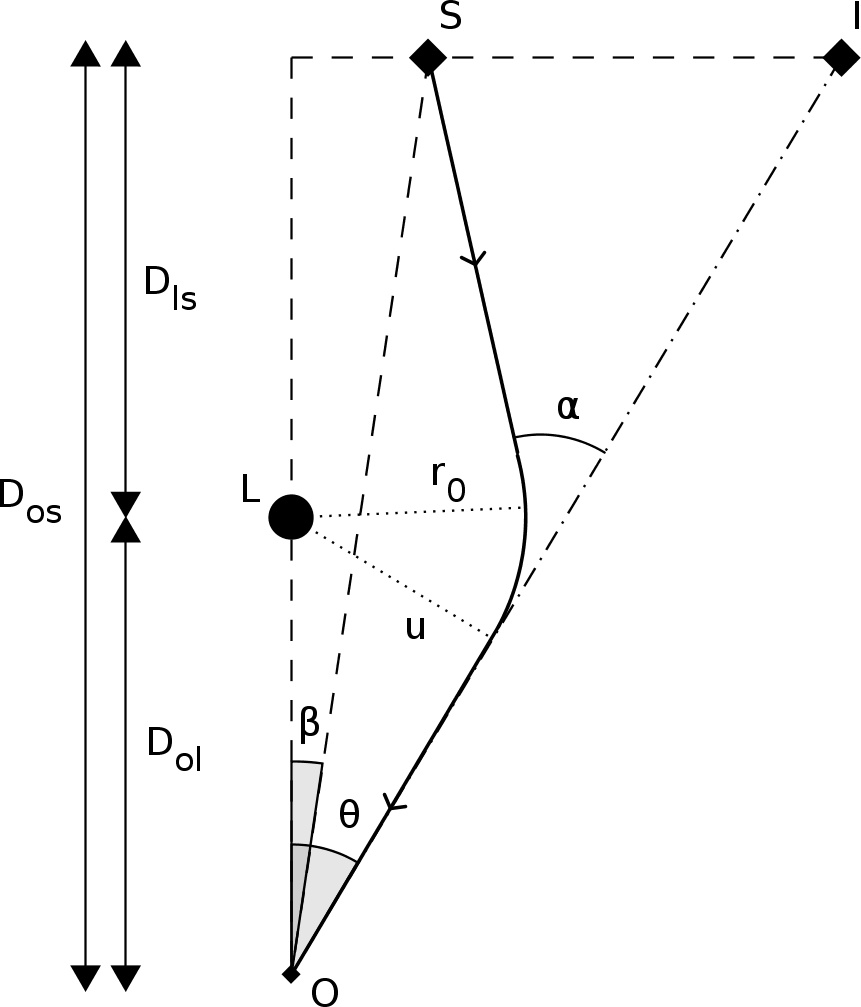}
\caption{Gravitational lensing setup. The distance $r_{0}$ is  the closest approximation of the light rays to the lens, and $u$ denotes the impact parameter.}
\end{figure}
\begin{equation}\label{abr}
ds^{2} = A(r) dt^{2} - B(r)dr^{2} - r^{2} d\Omega^{2},
\end{equation} 
yields
\begin{equation}
\alpha(r) = \int_{r}^{\infty} \frac{B(r)^{1/2}}{r}\left[\left(\frac{r}{r_{0}}\right)^{2}\left(\frac{A(r_{0})}{A(r)}\right) - 1\right]^{1/2}   {dr}.
\end{equation}  
In the weak field regime, corresponding to the light passing far way from the compact object event horizon, the angles in Eq. $\left(8\right)$ can be expanded to the first order. In the particular scenario of the Schwarzschild solution, the deflection reads $
\alpha (r_{0}) = \frac{4GM}{c^{2}r_{0}},$ 
and two antipode images are observed. The separation between these two images is the well-known Einstein radius
$\theta_{E} = \frac2c\sqrt{\frac{GMD_{ls}}{D_{os}D_{ol}}}.$ 
When the lens and the source are aligned, $\beta = 0$, a ring of radius $\theta_{E}$, known as the Einstein ring, is formed around the lens. {}{Moreover, typical distances between images in gravitational lensing are of the order of the Einstein radius.} 

Here we are interested in the full non-linear regime of GR, that translates in considering the light trajectory near the event horizon. In this setup, the deflection angle increases and, for some impact parameter, becomes larger than $2\pi$, meaning a photon looping around the black hole, before travelling towards the observer. The images originated from those photons are known as relativistic images \cite{Virbhadra:1999nm}. If one continues to decrease the impact parameter, the number of loops shall increase until $\alpha$ diverges, and the photon crosses the event horizon. Therefore, instead of {}{an} image at each side of {}{the} lens, that appear in the weak field regime, the observer {}{realises an} infinite sequence of images, {}{at} each side of the lens \cite{Bozza:2001xd}. The so-called \textit{photon sphere}, $r_{m}$, defining unstable orbits around the black hole \cite{Whisker:2004gq}, is the largest solution of the equation 
$
(\ln A(r))^\prime = {1}/{r},$
and the correspondent impact parameter reads $
u_{m} = {r_{m}}/{A(r_{m})}.$ 
As the closest approximation $r_{0}$ {}{tends} to the photon sphere, the deflection angle diverges logarithmically \cite{Bozza:2002zj}. Doing the appropriate expansion around $r_m$, the expression for the deflection angle, in the {}{so-called} strong deflection limit (SDL), reads \cite{Bozza:2002zj}  
\begin{equation}
\alpha (\theta) = -\bar{a} \ln\left(\frac{ \theta D_{ol} }{ u_{m} } - 1\right) + \bar{b} + \mathcal{O} (u - u_{m}), \label{deflectionlog}
\end{equation}
where $\mathcal{O} (u - u_{m})$ represents higher order terms, and $\bar{a}$, $\bar{b}$ are the SDL coefficients that  depend upon $r$, $A(r)$, and $B(r)$, calculated at the point $r_{m}$:
\begin{eqnarray}
\bar{a} &=& \frac{{\rm R}(0,r_{m})}{2\sqrt{\mathfrak{b}}},\label{aaa} \\
\bar{b} &=& \bar{a} \ln\left(\frac{2\mathfrak{b}}{A(r_{m})}\right) + b_{{\rm R}} - \pi,\label{bbb}
\end{eqnarray}
where
\begin{eqnarray}
\!\!\!\!\!\!{\rm R}(z,r_{0}) &=& \frac{2\sqrt{B(r)A(r)}}{rA^{\prime}(r)}\left(1 - A(r_{0})\right), \\
\!\!\!\!\!\!z &=& \frac{A(r) - A(r_{0})}{1 - A(r_{0})},\label{zzzz}\\ 
\!\!\!b_{R}\!\!\!&=\!&\!\!\!\!\!  \int_{0}^{1}\!\!\!\left[{\rm R}(z,r_{m})f(z,r_{m})\!-\!{\rm R}(0,r_{m})f_{0}(z,r_{m})\right]dz, \\
\mathfrak{b} &=& \frac{(1 - A(r))^{2}}{2r^{4}A^{\prime 3}(r)}\left[-  2r^{3}A^{\prime \prime}(r)A(r)\right.\\ \nonumber 
&&\left. +2r\left(1-4r\right)A^{\prime}(r)A(r) 
+4r^3A^{\prime }(r)^2\right]\Big\vert_{r=r_0},
\end{eqnarray}
and
\begin{eqnarray}
\!\!\!f(z,r_{0}) \!&=& \!\!\left(\!A(r_{0}) \!-\! \left[\left(1\!-\!A(r_{0})\right)z \!-\! A(r_{0})\right]\frac{r_{0}^{2}}{r^{2}}\right)^{-1/2}\,, \\ 
f_{0}(z, r_{0}) &=& \left({\mathfrak{b} z^{2}}+\mathfrak{a} z \right)^{-1/2}\,,
\end{eqnarray}
being $\mathfrak{a}$ given by
\begin{equation}
\mathfrak{a} = \frac{1 -A(r)}{r^{2}A^{\prime}(r)}\left[2rA(r) - r^{2}A^{\prime}(r)\right]\Big\vert_{r=r_0}.
\end{equation}
In the above expression, all the functions of $r$ are evaluated at {}{$r_0$}, obtained by inverting Eq. (\ref{zzzz}) \cite{Bozza:2002zj}. The parameters $\bar{a}$ and $\bar{b}$ play a prominent role in measuring the angular difference from the outmost image and the adherent point related to the sequence of subsequent images. We shall better discuss the underlying physical properties of those parameters in the analysis of Sect. IV. 
Eq. \eqref{deflectionlog}, together with considerations regarding the angles $\theta$, $\beta$, and $\alpha$, allows us to derive the observable quantities for the gravitational lensing, which shall be accomplished in Sect.  IV.

In the next section we shall present the MGD and CFM metrics paradigm and physical motivation to, subsequently, study their gravitational lensing effects. In addition, we are going to analyse the modifications of this approach to the listed black holes in the literature \cite{Bozza:2001xd,Bozza:2002zj,Majumdar:2004mz}, and references therein.

\section{MGD and CFM metrics}

In this section we briefly present the main physical features  regarding the minimal geometric deformation procedure and the CFM solutions. 
\subsection{The MGD procedure}
\label{MGD}
In brane-world scenarios, the brane self-gravity is encrypted in the brane tension $\sigma$. The brane tension was previously bounded in  the DMPR (Dadhich-Maartens-Papadopoulos-Rezania)} \cite{dadhich} and CFM solutions, by the classical tests of GR~\cite{Bohmer:2009yx}.
In the MGD procedure context, the bound $\sigma \gtrsim  5.19\times10^6 \;{\rm MeV^4}$ was obtained \cite{Casadio:2015jva},  providing a  stronger bound, contrasted to the one provided by cosmological nucleosynthesis. 
The MGD underlying geometry encompasses  high-energy corrections, since $\sigma^{-1/2}$ plays the role of
the 5D fundamental length scale\footnote{The units
 $c=G=1$ shall be adopted, $G$ denoting the 4D Newton constant,
unless otherwise specified.}.
The MGD is a deformation of the Schwarzschild solution, and devises observational effects that are originated by the PPN 
parameter $\zeta\simeq (\sigma^{-1/2}/R)^2$, describing a 4D geometry  that surrounds a star of radius $R$, localised on the brane. Moreover, the MGD is naturally led to the Schwarzschild metric, when $\sigma^{-1}\to 0$. 
In order to proceed,  the Einstein effective brane equations read~\cite{GCGR}
\begin{equation}
\label{gmunu}
G_{\mu\nu}=-\Lambda\, g_{\mu\nu}
-\mathring{T}_{\mu\nu}.
\end{equation}
The effective energy-momentum tensor   
$\mathring{T}_{\mu\nu}
=
T_{\mu\nu}+{\it E}_{\mu\nu}+\frac{6}{\sigma}S_{\mu\nu}
$ splits into the brane matter energy-momentum tensor ($T_{\mu\nu}$), the   high-energy
Kaluza-Klein induced tensor ($S_{\mu\nu}$) and non-local corrections induced by the electric component of the Weyl tensor  (${E}_{\mu\nu}$). 
MGD metrics  are exact solutions of Eq. \eqref{gmunu} \cite{ovalle2007}, yielding physical stellar inner   solutions \cite{Ovalle:2007bn}, having Schwarzschild outer solution  that does not jet energy into
the extra dimension~\cite{darkstars}. 5D solutions were further obtained in various contexts ~\cite{bs1,Casadio:2013uma,daRocha:2013ki,anjos}.
\par The MGD, requiring that GR must be the low energy dominant regime
$\sigma^{-1}\to 0$, derives  a deformed radial component of the metric, by bulk effects, yielding~\cite{covalle2}
\begin{eqnarray}
\label{edlrwssg}
B(r)
&=&
\nu(r)
+f(r)
\ ,
\end{eqnarray}
where (hereon we denote $\frac{GM}{c^2}\mapsto M$) 
\bea\!\!\! f(r)\!&\!\!=\!\!&\!e^{-I}\!\left(\int_0^r\!\!\!\frac{e^I}{\frac{A'^{2}}{2A^2}\!+\!\frac{2}{x}}\!
\left[H(p,\rho,A)\!+\!\frac{\rho^2\!+\!3\rho\,p}{\sigma}\right]
\!dx\!+\!\zeta\right),\nonumber\\
 I(r)&=&
\label{I}
\int^r_{r_0}
\frac{\!\frac{A''A}{A'^{2}}\!-\!1
\!+\!\frac{A'^{2}}{A^2}\!+\!\frac{2A'}{Ar}\!+\!\frac{1}{r^{2}}}
{\frac{A'}{2A}+\frac{2}{r}}\,dr
\ ,\\
\nu(r)
&=&
1-\strut\displaystyle\frac{2\,\mathring{M}}{r}
\ ,
\end{eqnarray} where $\mathring{M}=M$, for $r>R$, or $\mathring{M}=m(r)$, for $r\,\leq\,R$, 
where $m(r)\equiv\frac{1}{2r}\!\!\int_0^r\! x^2\rho\, dx$, and  $M_0 = M\vert_{\sigma^{-1}\to0}$. 
The function $H$ in Eq.~\eqref{I}, \begin{eqnarray}
\label{H}
H(p,\rho,A(r))
&\equiv&
\,p-\left[ \frac{A'}{A}\left( \frac{B'}{2B}+\frac{1}{r}\right)+(\ln B -1)r^{-2}\right.
\nonumber
\\
&&+
\left.\ln B\!\left(\!\frac{A''A}{A'^{2}}\!-\!1
\!+\!\frac{A'^{2}}{A^2}\!+\!\frac{2A'}{Ar}\right)\!
\right]\,,
\end{eqnarray} encompasses anisotropic effects of bulk gravity, the pressure, and the density. 
The 
deformation $f(r)$ in Eq.~\eqref{edlrwssg} is minimal \cite{ovalle2007}, 
$
f^+(r)=
\left.f(r)\right|_{p=\rho=H=0}
=
\zeta\,e^{-I}.$ The outer radial component in Eq. \eqref{edlrwssg} is given by
\begin{eqnarray}
\label{g11vaccum}
B_+(r)
=
{1-\frac{2\,M}{r}}
+\zeta\,e^{-I}.
\end{eqnarray}
 The parameter $\zeta$
equals the 5D correction to the vacuum, evaluated at the
star surface. Hence the MGD parameter $\zeta$ encodes a Weyl fluid~\cite{Casadio:2015jva}. 
Matching conditions, for $r<R$, yield 
\begin{equation}
\label{genint}
ds^2
=
A_-(r)\,dt^2-\frac{dr^2}{1-\frac{2m(r)}{r}+f^-(r)}-r^2\,d\Omega^2
\,,
\end{equation}
where $f^-(r)$ satisfies Eq.~\eqref{edlrwssg} with $H=0$. For $r>R$,  the metric, following Eq.~\eqref{g11vaccum}, reads 
\begin{equation}
\label{genericext}
ds^2
=
A_+(r)\,dt^2-\frac{dr^2}{1-\frac{2M}{r}+f^+(r)}-r^2\,d\Omega^2
\ ,
\end{equation}
since the star is surrounded by a Weyl fluid \cite{ovalle2007,Casadio:2013uma}. 
The MGD function $f=f^+(r)$ has the form 
\begin{equation}
\label{hhh}
f^+(r)
=    
\,\left({1-\frac{2M}{r}}\right)\left({1-\frac{3M}{2\,r}}\right)^{-1}\,
\frac{\zeta\ell}{r}
\ ,
\end{equation}
where $\ell$ is a length given by
\begin{equation}
\label{L}
\ell
\equiv
R{\left(1-\frac{3M}{2R}\right)}{\left(1-\frac{2M}{R}\right)^{-1}}
\,.
\end{equation}
In this way, the deformed outer metric reads
\begin{subequations}
\ba
\label{nu}
\!\!\!\!\!\!A(r)
&=&
1-\frac{2\,M}{r}
\ ,
\\
\!\!\!\!\!\!B(r)
&=&
\left(1-\frac{2\,M}{r}\right)
\left[1+{\zeta}\left({1-\frac{3\,M}{2\,r}}\right)^{-1}\,\frac{\ell}{r}\right],
\label{mu}
\ea
\end{subequations}
enclosing the vacuum solution in Ref.~\cite{germ} in the particular case
when $\zeta\,\ell={k}/{\sigma}$, $k>0$. 
 The outer MGD
at the star surface 
must be negative, otherwise a negative pressure 
 for a solid crust would appear~\cite{darkstars}. 
The outer geometry, governed by Eqs.~\eqref{nu} and \eqref{mu}, has two event horizons:
\be
r_S = 2\,M
\quad
{\rm and}
\quad
r_2=\frac{3M}{2} - \zeta\,\ell
\ .
\ee
However, one must have $r_2< r_S$, since the approximation $\zeta\sim \sigma^{-1}$ should hold in the GR limit.
This implies that the outer horizon  is the Schwarzschild one $r_S = 2\,M$.
It is worth noting the specific value $\zeta=-M/2$ would produce a
single horizon $r_S = 2\,M$. However, the limit $\sigma^{-1}\to 0$ 
does not reproduce the Schwarzschild solution, since $M_0 = M\vert_{\sigma^{-1}\to0}$.
On the other hand, 
the deformed event 
horizon  $r_S=2\,M$ is smaller than the Schwarzschild horizon 
$r_S=2\,M_0$. Hence 5D effects weaken the 
strength of the gravitational field, produced by the self-gravitating compact star.
\par

In particular, $\zeta$ can be derived by considering the exact inner brane-world solution
of Ref.~\cite{ovalle2007}.
\begin{eqnarray}
\label{betasigma}
\zeta(\sigma,R)
=
-\frac{C_0}{R^2\,\sigma}
\ .
\label{c0}
\end{eqnarray}
where $C_0\simeq 1.35$. 
The outer geometric deformation is finally obtained by using Eq.~\eqref{c0}
in Eq.~\eqref{hhh}, leading to
\begin{equation}
\label{hhhh}
\!f^+(r)
\!=\!
-\frac{C_0\ell_0}{R^2\,\sigma r}\,\left({1\!-\!\frac{2M_0}{r}}\!\right)\left({1\!-\!\frac{3M_0}{2\,r}}\right)^{-1}\!\!\!\!\!\!\!+{\cal O}(\sigma^{-2})
\ ,
\end{equation}
where $\ell_0=\ell(M_0)$, regarding Eq.~(\ref{L}). 
In addition, 5D effects are maximal at the star surface, and more perceivable for very compact stellar distributions.
Hence, the more compact the star, the larger $|\zeta|$ is.

{It is worth to mention that the brane tension $\sigma$, with special focus on Eqs. (\ref{c0}) and (\ref{hhhh}), should vary along the cosmological evolution of the Universe, since it is a scalar field \cite{CORDAS} or an intrinsic brane feature \cite{Gergely:2008fw,HoffdaSilva:2011bd}. 
{Along each defined phase of the Universe evolution -- except in the phase transitions -- the changes in the brane tension are tiny, being perceptible just across cosmological time scales. Around intervals of $10^7\sim 10^8$ years, changes in the brane tension are not so easy to detect, since one of the celebrated models regards  the E\"otv\"os brane models, wherein the brane tension is proportional to the Universe temperature (see Refs. [39, 40, 41]). Our analysis here are made in such context. Moreover, this analysis is useful to detect nowadays signatures provided by black holes that are  $10^7\sim 10^8$ (or more) light-years distant, which made it more realistic, due to this paradigm.}
{
\subsection{Casadio-Fabbri-Mazzacurati brane-world solutions}
\label{S2}
{The effective Einstein field equations on the brane have solutions that} are not solely governed by brane energy-momentum tensor. In fact, bulk gravity may impel a Weyl term on  the brane. Such kind of solutions are represented by the CFM metrics~\cite{Casadio:2002uv}, that are vacuum brane solutions, with PPN parameter $\delta$.
The Nordtvedt effect  provides the bound $|\delta-1|\lesssim0.00023$, whereas the observation of the deflection of light yields 
$|\delta-1|\lesssim 0.003$~\cite{WILL}.
The CFM metrics were derived when the constraint $A(r)=B^{-1}(r)$ is relaxed, in Eq. (\ref{abr}).
The CFM I solution is obtained by fixing $A(r)$ to be equal to the analogous  Schwarzschild metric coefficient and, subsequently, deriving 
$B(r)$ from the Einstein field equations \cite{cfabbri}. On the other hand, to derive the CFM II solution, the Reissner-Nordstr\"om type coefficient $A(r)$,  representing a tidal charge, is employed ~\cite{dadhich}. 
For the CFM I solutions, the metric coefficients in Eq.~\eqref{abr} are 
\beq
\!\!\!\!\!\!\!\!\!\!\!\!\!\!\!\!\!\!A_I(r)
&=&
1- \frac{2\,\GN\, M}{r},
\\
\qquad
\!\!\!\!\!\!\!\!\!\!\!\!\!\!\!\!\!\!B_I(r)
&=&
\frac{1-\frac{3\,\GN\,M}{2\,r}}
{\left(1- \frac{2\,\GN\,M}{r}\right)\left[1-\frac{\GN\,M}{2\,r}(4\delta-1)\right]}
\equiv
B(r).
\label{ar}
\eeq 
The event horizon $r={\rm R}$ on the brane is, hence, derived by the algebraic equation $1/B({\rm R})=0$. The sign of $(\delta - 1)$ implies that the corresponding black hole is either {hotter or colder} than the Schwarzschild black hole~\cite{cfabbri}. The physical singularities are governed by the singular points of curvature (Kretschmann) scalars. In fact,  the coordinates $r = 0$ and $r = {3\,\GN\,M}/{2}<{\rm R}_S$ are physical singularities, wherein the scalar $R_{\mu\nu\rho\sigma} R^{\mu\nu\rho\sigma}$ diverges.
 Moreover, the CFM II solution regards ~\cite{cfabbri,Casadio:2002uv}
\begin{eqnarray}
\!\!\!\!\!\!\!\!A_{II}(r)
&=&
1- \frac{2(2\delta-1)\,\GN\,M}{r}\\
\quad
\!\!\!\!\!\!\!\! B_{II}(r)\!
 &\!=\!&\!
 \frac{1}{(2\delta\!-\!1)^2}\!\left(2(\delta\!-\!1)\!+\!\sqrt{1\!-\! \frac{\!2(2\delta\!-\!1)\,\GN\,M}{r}}\right)^2
\ .
\label{ar112} 
\end{eqnarray}
A detailed study on the causal structure and the 5D black strings associated to both CFM
solutions can be found in Refs.~\cite{cfabbri,Casadio:2002uv,daRocha:2013ki}. The modified version of the  CFM solutions in the SDL regime appeared previously, in a different context, in Ref. \cite{Nandi:2006ds}.

In the next section we shall apply and analyse the above described metric to study the modifications on the gravitational lensing 
effects, when compared to other already obtained solutions. In particular, we shall investigate the role of the PPN parameter and the 
 brane tension in the MGD, on the observables derived  in the strong deflection limit.
 
\section{Observables in the {strong deflection limit}}

One of the important aspects of the SDL approach is the potential to identify different black hole solutions. However, the applicability of the approach is hugely increased by the possibility of testing extensions of GR. In fact, any testable features of  theories beyond GR could provide us important hints on the nature of gravity, in such regimes. Our aim, in this section, is to identify the deviation of the MGD and CFM solutions from the classical Schwarzschild black hole. It is accomplished by analysing the observables of the SDL found in Ref. \cite{Bozza:2002zj} as well as the time delay of relativistic images \cite{Bozza:2003cp} and, subsequently, comparing them with the standard Schwarzschild one.

Some observable features of the SDL regime can be calculated, by using only the expansion coefficients, namely, $\bar{a}$, $\bar{b}$, respectively in Eqs. (\ref{aaa}) and (\ref{bbb}), and $u_{m} = {r_{m}}/{A(r_{m})}$  \cite{Bozza:2002zj}. Fig.  \ref{fig_coeff_mgd} shows their  behaviour, upon varying the parameter $\zeta$ of the MGD solution. Figs. \ref{fig_coeff_CFM I} and \ref{fig_coeff_CFM II} show that the coefficients do not have an appreciable variation, in the allowed range of the parameter $\delta$, for the CFM I and CFM II solutions, as showed in Table II. It is worth to mention that the lines in Figs. 3 and 4 are not really straight, but just 
a resolutional consequence of the tiny range determined by the (currently observed) PPN parameter $\delta$. Furthermore, Figs. \ref{fig_coeff_mgd}, \ref{fig_coeff_CFM I} and \ref{fig_coeff_CFM II} show that the SDL coefficients are smooth in the allowed range.

\begin{figure}[!htb]
\centering
\includegraphics[scale=.67]{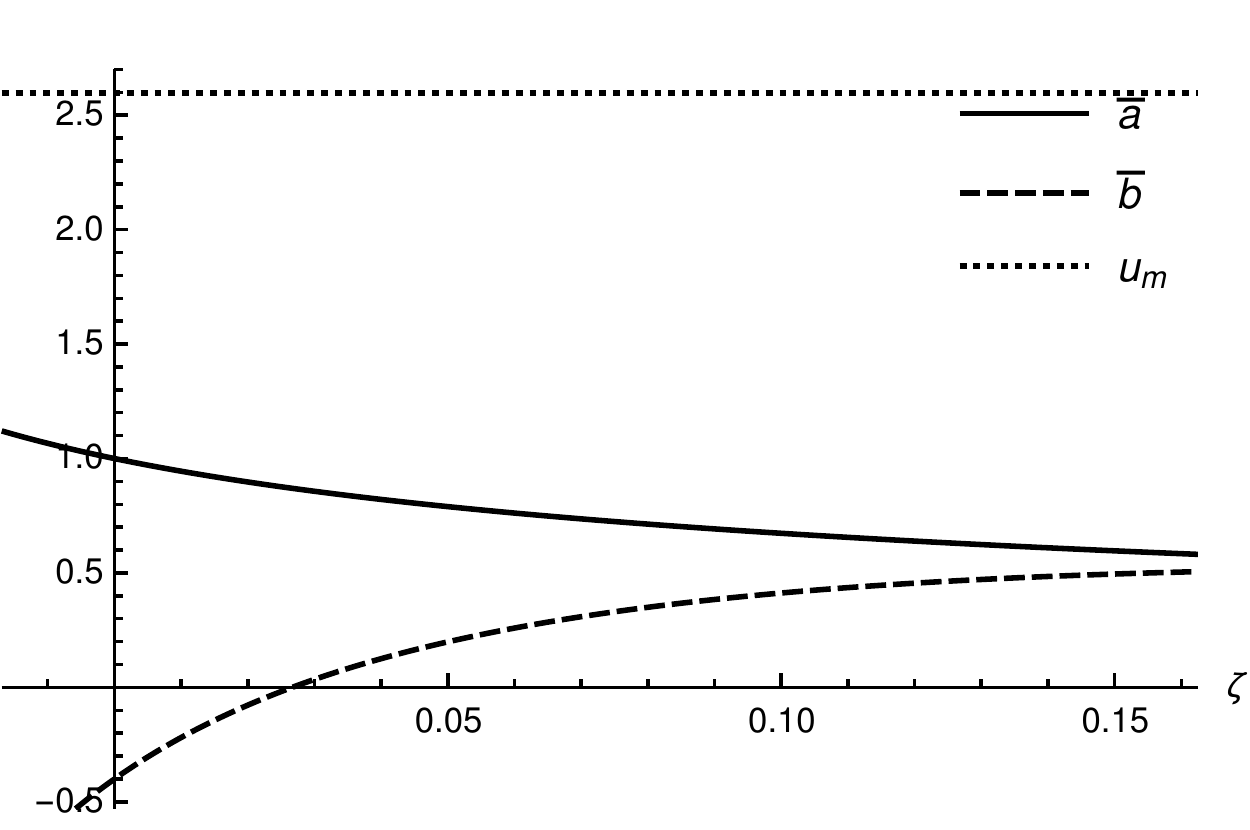}
\caption{SDL coefficients for the MGD solution as functions of the parameter $\zeta$ in Eq. (\ref{c0}).}
\label{fig_coeff_mgd}
\end{figure}

\begin{figure}[!htb]
\centering
\includegraphics[scale=.67]{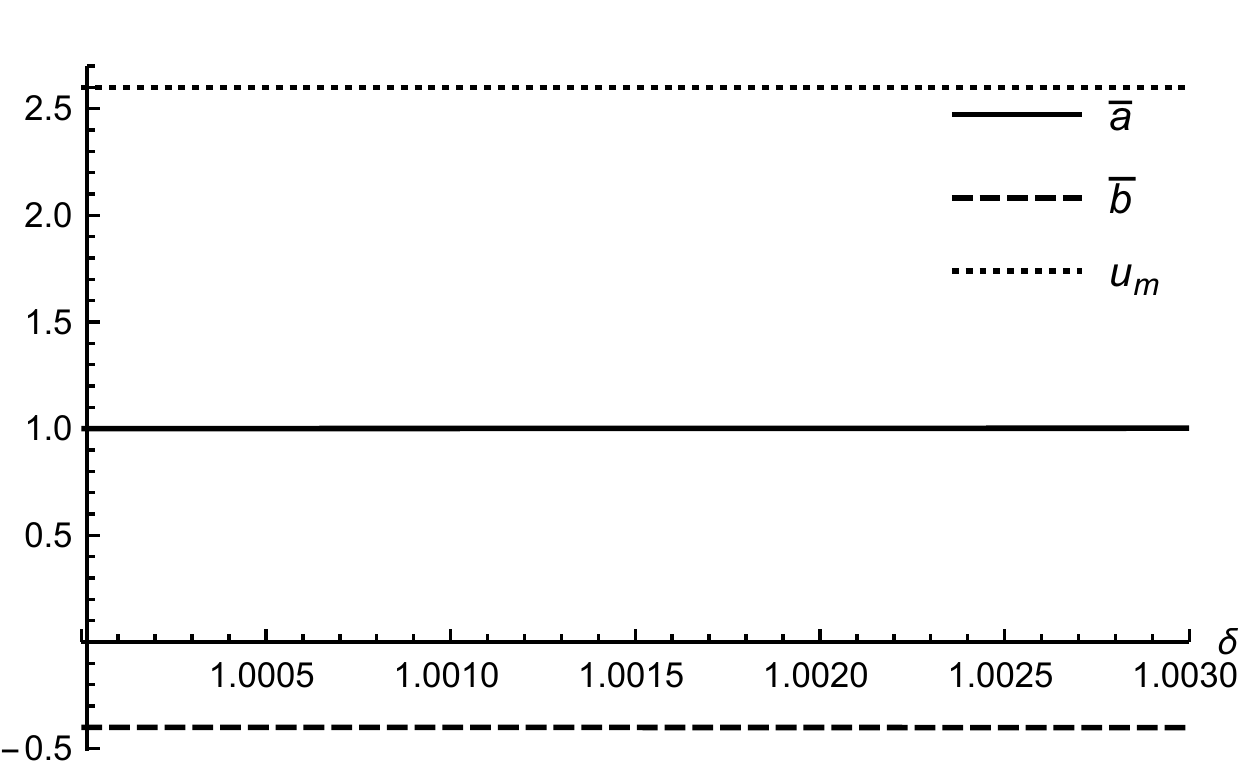}
\caption{SDL coefficients for the CFM I solution, as functions of the PPN parameter $\delta$.}
\label{fig_coeff_CFM I}
\end{figure}

\begin{figure}[!htb]
\centering
\includegraphics[scale=.67]{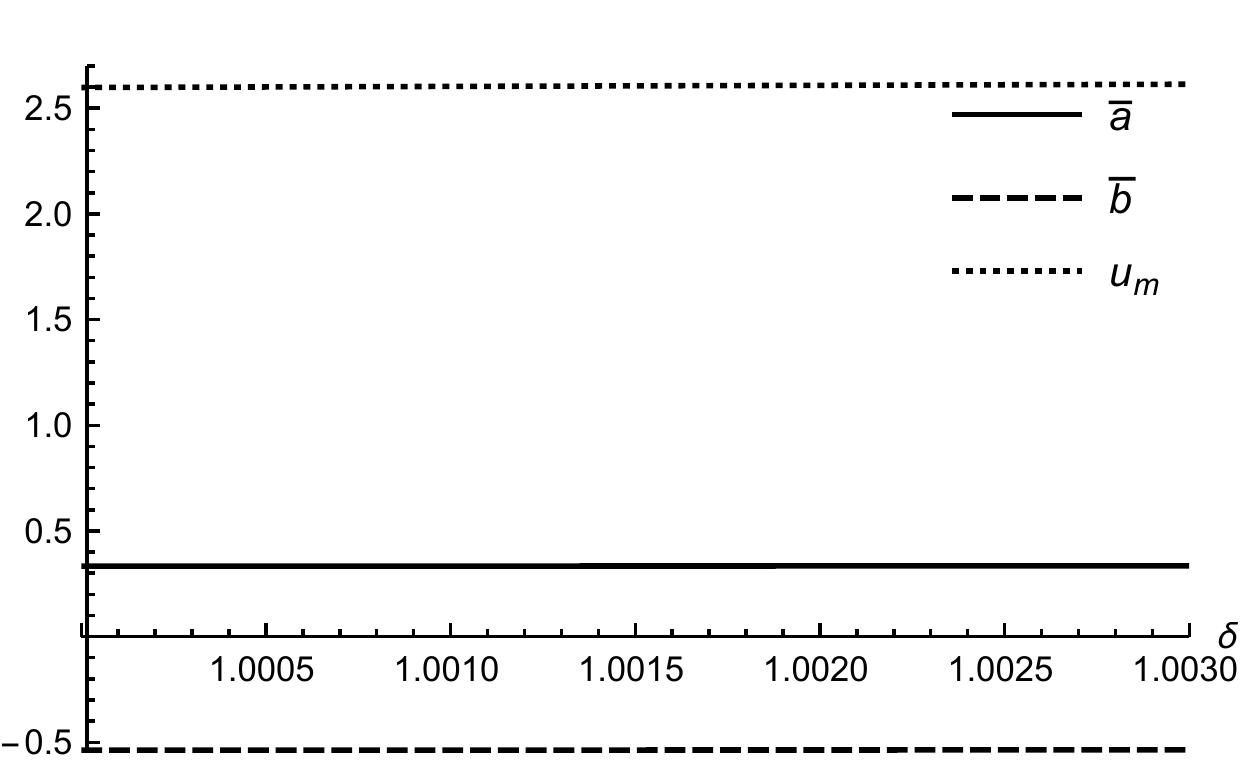}
\caption{SDL coefficients for the CFM II solution, as functions of the PPN parameter $\delta$.}
\label{fig_coeff_CFM II}
\end{figure}

The first observable introduced in \cite{Bozza:2002zj} is the angular position $\theta_\infty$ of the accumulating relativistic images. The second, denoted by $s$, is related to the distance of relativistic images. It is defined as the angular distance between the largest and smallest orbit of the light rays winding around the black hole. The third one is the magnification of the images after of the lensing effect. The details of the derivation of those observables are not very involved, however here we are going to only sketch those derivations, referring to the original paper for more details.

Despite of the angle $\theta$ between the lens and the image being large in the SDL, the effects are more prominent when the source, the lens, and the  observer are highly aligned \cite{Virbhadra:1999nm}. It means that the deflection angle $\alpha$ should be a small deviation of $2n\pi$, that is $\alpha(\theta)=2n\pi+\Delta \alpha_n$, with $\Delta \alpha_n$ small. The next step is to find the angles $\theta_n^0$ such that $\alpha(\theta_n^0)=2n\pi$, which represents a deviation $\Delta \alpha_n$, and expand $\alpha(\theta)$ around $\theta_n^0$. Hence, the lens equation \eqref{deflectionlog} straightforwardly yields  
\begin{equation}
\theta_n^0=\frac{u_m}{D_{ol}}\left[1+\exp\left(\frac{\bar{b}-2n\pi}{\bar{a}}\right)\right]\,,
\end{equation}
resulting in
\begin{equation}
\Delta\alpha_n=\frac{\bar{a}D_{ol}}{u_m \exp\left(\frac{\bar{b}-2n\pi}{\bar{a}}\right) }(\theta_n^0-\theta).
\end{equation}
It allows us to define the two following observables,\vspace*{-0.2cm}
\begin{align}
\theta_\infty\equiv&\lim_{n\rightarrow \infty}\theta_n^0=\frac{u_m}{D_{ol}},\\
s\equiv &\; \theta_1-\theta_\infty=\theta_\infty \exp\left(\frac{\bar{b}-2\pi}{\bar{a}}\right).
\end{align}\vspace*{-0.03cm}
\!\!The magnification resulting from the lensing effect, defined as the ratio of the flux of the image to the flux of the unlensed source \cite{Virbhadra:1999nm}, is the third observable quantity in the SDL regime. Given the flux\vspace{-0.3cm}
\begin{equation}\vspace*{-0.2cm}
\mu=\frac{\sin \theta}{\sin \beta}\left(\frac{d \beta}{d \theta}\right)^{-1},
\end{equation}
the ratio of flux of the images to the flux of the source, expressed as function of the SDL coefficients, reads \cite{Bozza:2002zj,Bozza:2009yw}\vspace*{-0.6cm}
\begin{equation}
k=\frac{\mu_1}{\sum_{n=2}^\infty \mu_n}=\exp\left(\frac{2\pi}{\bar{a}}\right).\end{equation}

Another useful information that can be extracted from the SFL regime and its coefficients is the time delay of relativistic images. A simple formula was derived by Bozza and Mancini in \cite{Bozza:2003cp}, relating the delay between the $n$-loop image and the $m$-loop image. According to their result, when the  lens and the observer are nearly aligned and the black hole has spherical symmetry,  the time delay is given by
\begin{equation}
\Delta T_{n,m}=T_{n,m}^0+T_{n,m}^1\,,\end{equation}
where
\begin{equation}
\Delta T_{n,m}^0 = 2\pi(n-m)u_m\,,\end{equation}
and the first correction  is given by
\begin{align}\nonumber
\Delta T_{n,m}^1=&2 \sqrt{\frac{B(r_m)}{A(r_m)}} \sqrt{\frac{u_m}{\hat{c}}} \exp\left(\frac{\bar{b}}{2\bar{a}}\right) \times\\
&\left[\exp\left(-\frac{\pi  m}{\bar{a}}\right)-\exp\left(-\frac{\pi  n}{\bar{a}}\right)\right]
\end{align}
with
\begin{equation}
\hat{c}=\frac{1}{4\sqrt{A(r_m)^3C(r_m)}}\left[A(r_m)C''(r_m)-A''(r_m)C(r_m)\right].\end{equation}
The first correction $T_{2,1}^1$, in the Schwarzchild case, was shown to contribute with only 1.4\% {}{to} the total time delay \cite{Bozza:2003cp}. 
 The cases analysed in the present paper lead to a correction of order of $0.2-2$\%, 1.5\% and 0.0005\%, for the MGD, CFMI, and CFMII {}{metrics}, respectively.

{}{Having} introduced all the observable quantities, now we can use the known data from Sagittarius A$^*$, the galactic supermassive black hole at the center of our galaxy, to compare SDL lensing results of the Schwarzschild, MGD, and CFM solution. The observables and the SDL coefficients for MGD and CFM solutions are displayed in the Tables I \& II. In both cases we calculated the results considering the allowed range of the parameter $\zeta$ for the MGD metric and the $\delta$ for the CFM metrics. We have used the recent results for the mass and distance of the Sagittarius A$^*$, updating also the results for the Schwarzchild black hole. According to \cite{SgtA}, they are given by $M=4.02\times 10^6M_\odot$ and $D_{ol}=7.86$ kpc respectively. In what follows, the value $R=1.437 R_S$, regarding the MGD solutions, shall be taken into account in $\zeta$, corresponding to a compact object surface, by considering  the current value for the cosmological constant in the matching conditions at the {}{compact object} surface. In addition, in the tables below we denote, as usual, by $s=\theta_1-\theta_\infty$ the angular difference from the outmost image and the adherent point formed by the others for the limit $n\to\infty$; $k_m=2.5\log k$ denotes the magnitude; $u_m/R_S$ is the normalised minimum impact parameter; $\bar{a}$ and $\bar{b}$ denote strong deflection limit coefficients that constitute the deflection angle; $\Delta T_{2,1}$ denotes the time delay between the 1-loop and 2-loop relativistic images given in Schwarzchild time ($2GM/c^3 \approx 39,56s$) and $\Delta T_{2,1}(\mbox{min})$ is the same time delay, given in minutes. 
\begin{widetext}
\begin{center}
\begin{tabular}{||c||c||c|c|c|c|c|c||}
\hline\hline 
 & Schwarzschild& \multicolumn{6}{c|}{MGD} \\ 
\cline{3-8}\cline{3-8} 
 &  ($\zeta\equiv0$)&$\zeta= -1.69\times 10^{-2}$ & 1.31$\times 10^{-2}$ & 4.31$\times 10^{-2}$ & 7.31$\times 10^{-2}$ & 10.31$\times 10^{-2}$ & 13.31$\times 10^{-2}$ \\
\hline\hline 
$\theta_\infty$ $(\mu{\rm arcsec})$ & 20.21 & 26.21 &26.21 & 26.21 &26.21 & 26.21 & 26.21\\\hline 
$s$ $(\mu{\rm arcsec})$& 0.0328 & 0.0291 &0.0452 & 0.0136 & 0.0074 & 0.0040 & 0.0022 \\ \hline 
$k_m$ & 6.82 & 6.09 & 7.34 & 8.41 & 9.36 & 10.22 & 11.00 \\ \hline 
$u_m/R_S$ & 2.6 & 2.6 & 2.6 &2.6 &2.6 & 2.6 & 2.6 \\ \hline 
$\bar{a}$ & 1 & 1.12 & 0.93 & 0.81 & 0.73 & 0.67 & 0.62 \\ \hline 
$\bar{b}$ & $-$0.4002 & $-$0.8459 & $-$0.1715 & 0.1499 & 0.3225 & 0.4203 & 0.4759 \\  \hline 
$\Delta T_{2,1}$ &16.573&16.646 &16.529&16.457&16.413&16.385&16.366 \\ \hline
$\Delta T_{2,1}(\mbox{min})$&10.93 &10.97&10.90 &10.85&10.82&10.80&10.79 \\ 
\hline\hline
\end{tabular}\medbreak
Table I. Observables in the strong deflection limit for the MGD solution.  
\end{center}

\begin{center}
\begin{tabular}{||c||c||c|c|c||c|c|c||}
\hline\hline 
 & Schwarzschild &   \multicolumn{3}{c|}{CFM I}&\multicolumn{3}{c|}{CFM II} \\ 
\cline{3-8} \cline{3-8}
 & ($\delta\equiv1$) &  $\delta=1.001$ &  $1.002$ &  $1.003$ &  $\delta=1.001$ &  $1.002$ &  $1.003$  \\\hline 
\hline 
$\theta_\infty$ $(\mu{\rm arcsec})$& 26.21 & 26.21 & 26.21 &26.21 &  26.26 & 26.31 & 26.36  \\\hline 
$s$ $(\mu{\rm arcsec})$& 0.0328 & 0.0329 &0.0331 & 0.0332  & $\sim 10^{-8}$ & $\sim 10^{-8}$ & $\sim 10^{-8}$ \\\hline 
$k_m$ & 6.82 & 6.817 & 6.813 & 6.808  & 20.436 & 20.410 & 20.377  \\\hline 
$u_m/R_S$ & 2.6 & 2.6 & 2.6 & 2.6&2.6 & 2.61& 2.61 \\\hline 
$\bar{a}$ & 1 & 1.0007 & 1.0013& 1.0020  & 0.3338 & 0.3348 & 0.3348  \\\hline 
$\bar{b}$ & $-0.4002$ & $-$0.4007 & $-$0.4011 & $-$0.4016  & $-$0.5368 & $-$0.5364 & $-$0.5361  \\ \hline
$\Delta T_{2,1}$ &16.573 &16.574&16.574&16.575&16.357&16.390&16.422  \\\hline
$\Delta T_{2,1}(\mbox{min})$ &10.93 &10.93&10.93&10.93&10.78&10.81&10.83  \\
\hline 
\hline 
\end{tabular}\medbreak
{Table II. Observables in the strong deflection limit for both the CFM solutions. } 
\end{center}
\end{widetext}

\section{Concluding remarks and outlook}

 The effects of gravitational lensing in the strong field regime are concretely noticeable   wherein the thorough capture of the photon by the black hole is regarded. The so-called inversion problem is capable to uniquely determine the nature of the black hole, from the parameters associated with the successive images formation, as  their relative distances, the asymptotic position approached by the set of images,  their measured positions  and the time delay of relativistic images, accordingly. 
Regarding both the CFM solutions, and the MGD of GR, we calculated the observables of the gravitation lensing of a background source in the strong field regime. The  angular difference  $s$ from the outmost image and the adherent point formed by the other images; the parameter $k_m$ that reveals the image magnification; the normalised minimum impact parameter $u_m/R_S$ were studied, unraveling a precise observable signature 
for the CFM I, the CFM II, and the MGD solutions. Moreover, the delay between relativistic images is significant for MGD and CFMII solutions, being a potentially key information on the caracterisation of the solution that may model the supermassive  black hole in the centre of our galaxy. 
Based upon the  satellite mission of the ESA of astrometric measurements, the accuracy can reach 7 $\mu$arcsec \cite{Jordan:2008ky}, the typical signature of CFM I, CFM II, and MGD solutions are, mainly, observable by the parameters $k_m$ and $\theta_\infty$, as well as the time delay. It is worth to mention that the CFM II solution is hugely different, when compared to the Schwarzschild, CFM I, and MGD solutions, in what regards the parameter the magnitude $k_m$. This analysis was presented and encrypted in Tables I -- for different values of the brane tension, according to Eq. (\ref{c0}), and II (for different values of the PPN parameter). The signatures of the MGD, on the other hand, could be evinced by the combinations of the parameters $k_m$ and $s$. However, even the difference on the values of $s$ for the MGD and Schwarzschild being potentially higher than one order of magnitude, it is still a little beyond the current ESA resolution. Thus, from the observational perspective, the astrophysical data should be better improved in order to apply the thorough potential of the gravitational lensing in the strong field regime.

\subsection*{Acknowledgements}

RTC and AGS~are supported by CAPES.
RdR~is grateful to CNPq (grants No. 303293/2015-2 and No.~473326/2013-2),
and to FAPESP (grant No.~2015/10270-0), for partial financial support. 

\bibliography{bib_lensing}{}

\end{document}